\documentclass[onecolumn,showpacs,prb,superscriptaddress]{revtex4}

\usepackage{color}
\usepackage{tabularx}
\usepackage{epsfig}
\usepackage{graphicx}

\begin{document}

\title{Heisenberg Spin Glass Experiments and the Chiral Ordering Scenario}

\author{I.~A.~Campbell}
\affiliation{Laboratoire des Collo\"ides, Verres et Nanomat\'eriaux,
Universit\'e Montpellier II, 34095 Montpellier, France}

\author{Doroth\'ee C.~M.~C.~Petit}
\affiliation{Blackett Physics Laboratory, Imperial College London, Prince Consort Road, London SW7 2BW, UK}

\begin{abstract}

An overview is given of experimental data on Heisenberg Spin Glass materials
so as to make detailed comparisons with numerical results on
model Heisenberg spin glasses, with particular reference to the chiral driven
ordering transition scenario due to Kawamura
and collaborators. On weak anisotropy systems, experiments show
critical exponents which are very similar to those
estimated numerically for the model Heisenberg chiral ordering transition
but which are quite different from those at Ising spin glass transitions.
Again on weak anisotropy Heisenberg spin glasses, experimental torque data show well defined in-field transverse ordering transitions up to strong applied fields, in contrast to Ising spin glasses where fields destroy ordering.
When samples with stronger anisotropies are studied,
critical and in-field behavior tend progressively towards the Ising limit.
It can be concluded that the essential physics of laboratory Heisenberg spin glasses mirrors that of model Heisenberg spin glasses, where chiral ordering has been demonstrated numerically.\\


\end{abstract}


\maketitle

\newpage

\section{Introduction}

The Spin Glass phenomenon was discovered in 1972 by Canella and Mydosh \cite{canella:72}$^)$. While  carrying out standard  systematic measurements on dilute alloys of magnetic elements in non-magnetic metal hosts, they made the remarkable and entirely unexpected observation that there is a sharp cusp in the variation with temperature of the low field linear ac susceptibility $\chi(T)$ of dilute ${\bf Au}Fe$ alloys. The  existence of this cusp implies that there is a hidden but well defined transition in this dilute alloy, although the magnetic impurities are distributed at random and interactions between spins are complex. Careful experimental work on a number of dilute magnetic alloys have shown that the  non-linear susceptibility (with terms proportional to $H^3$, $H^5$ etc.) diverges at the cusp temperature $T_c$, and that critical exponents can be defined and  measured, confirming the existence of a {\it bona fide} phase transition of an entirely new type. Numerous materials have now been shown to be spin glasses, and their properties have been studied extensively. Unusual dynamic and "memory" effects have attracted particular attention.

In 1975 an astute theoretical model was introduced by Edwards and Anderson (EA) ~\cite{edwards:75}$^)$; spins are placed on all sites of a regular lattice but the interactions between near neighbor spins are chosen to be random, with arbitrary sign. The EA spin glass order parameter is defined as $q_{SG}=\lim_{t\rightarrow\infty}<S_i(0)\cdot S_i(t)>$.
Model spin glasses,  mainly of the EA family, have been the subject of thousands of theoretical and numerical  studies. Because of technical and conceptual advantages most of the work has been carried out on Ising systems; the mean field (infinite dimension) Sherrington-Kirkpatrick (SK) \cite{sherrington:75}$^)$ version of the model has a full Replica Symmetry Breaking (RSB) solution \cite{parisi:83}$^)$. Numerical simulations show clearly that the EA Ising models in dimension 2 do not order at temperatures above zero but that there are finite SG ordering temperatures for dimensions 3 and above. In despite of the deceptive simplicity of the model, more than thirty years after its introduction important basic questions still remain to be resolved for the Ising EA SG model in finite dimensions.
Theoretical attention has been focused on a long-standing controversy as to the correct description of the physics of the ISG state below $T_c$ : does it resemble
the RSB solution of the mean field model,  or has it a much simpler structure described by the scaling or "droplet" approach \cite{fisher:88,moore:98}$^)$ ? In-field properties should distinguish between RSB and scaling models, because for RSB there should be a phase transition line under an applied magnetic field, while for the scaling model the true phase transition exists only in zero field, and any apparent transition line under field is a transitory relaxation effect. Experimental and numerical evidence is now predominantly in favor of the latter
interpretation \cite{mattsson:95,jorg:08,katzgraber:09}$^)$.

The original spin glasses however consisted of alloys containing three component vectorial localized spins and so are Heisenberg rather than Ising, with the magnetic impurities distributed at random in a non-magnetic metallic host. Typical alloys of this class are  ${\bf Au}Fe$, ${\bf Cu}Mn$, ${\bf Ag}Mn$, ${\bf Au}Mn$, ${\bf Au}Cr$, ... with impurity concentrations of the order of a few percent. Experimentally these systems clearly have well defined spin glass transition temperatures of a few degrees per percent impurity. Paradoxically pioneering numerical simulations on a Heisenberg version of the 3d EA model indicated zero temperature transitions \cite{banavar:82,mcmillan:85,olive:86}$^)$, raising the question as to the adequacy of this short range interaction model for the real-life physical systems where interactions are RKKY-like and so long range. Alternatively, an entirely novel chirality driven ordering scenario was proposed by H. Kawamura and collaborators \cite{kawamura:92}$^)$ for the standard 3d Heisenberg EA model with short range spin-spin interactions. Chirality \cite{villain:79}$^)$ is a multispin variable representing the handedness of the non-collinear
structures induced by frustration, Ref.~ \cite{viet:09a}$^)$. For Heisenberg spins on an sc lattice, the local chirality at the $i$-th
site and along the $\mu$-th axis $\chi_{i\mu}$ is defined for three
neighboring Heisenberg spins by a scalar
\begin{equation}
\chi_{i\mu} = {\bf S}_{i+{\bf r}} \cdot ({\bf S}_i \times {\bf S}_{i-{\bf r}})
\label{chirality}
\end{equation}
where ${\bf r}$ denotes a unit vector along the
$\mu$-th axis. For $L^3$ spins there are in total $3L^3$ local chiral variables. A chiral glass parameter $q_{CG}$ is defined in an analogous way to $q_{SG}$.

On the proposed chiral scenario, chiral glass order appears at a temperature well above the spin glass ordering temperature. There have been a number of claims that the chiral glass order does not set in at a higher temperature than the spin glass order in the HSGs \cite{lee:03,campos:06,lee:07,fernandez:09}$^)$. However successive interpretations of the numerical data are not always consistent with each other, and further analysis together with improved numerical work have shown some of these interpretations to be flawed \cite{campbell:07,viet:09a}$^)$. Direct $3d$ numerical results \cite{hukushima:00,hukushima:05,imagawa:04,viet:09,viet:09a}$^)$ reinforced by indirect arguments from $1d$ simulations \cite{matsuda:07}$^)$ provide evidence strongly in favor of the chiral driven ordering scenario for a pure short range spin-spin interaction 3d EA Heisenberg model. To summarize the conclusions drawn in Ref.~\cite{viet:09}$^)$ from the analysis of carefully equilibrated numerical data up to $L=32$ : when the chiral sector is fully taken into account there is decoupling between chiral and spin sectors above a certain  sample size (which simulations indicate to be $L \approx 12$); in the thermodynamic limit the chiral degrees of freedom order at a finite $T_{CG}$ which occurs at a temperature well above a spin $T_{SG}$ of the conventional (Ising-like) type. A negative dip of the chiral Binder cumulant at temperatures above $T_{CG}$ indicates that the chiral ordering has a 1-step RSB character. The chiral phase transition persists under very strong applied fields \cite{imagawa:04}$^)$ as the coupling between the field and the chiral glass order parameter is much weaker than between the field and the spin glass order parameter. This model is presented fully in a companion article by H. Kawamura \cite{kawamura:09}$^)$.

In all real Heisenberg spin glass materials in addition to the pure spin-spin exchange interactions there are anisotropic interaction terms, principally of the Dzyaloshinsky-Moriya (DM) \cite{dzyaloshinsky:58,moriya:60}$^)$ type; when weak anisotropy terms are introduced into the numerical model \cite{imagawa:04}$^)$ the spin sector becomes weakly re-coupled in a different manner to the chiral sector and the onset of chiral ordering can be expected to be "revealed" through the spin behavior.

This article will summarize some of the existing experimental information on real laboratory spin glasses. It will be seen that the properties of the laboratory Heisenberg spin glass materials show striking similarities to those of the numerical Heisenberg spin glasses as analysed on the basis of the chiral approach.

\section{Dilute magnetic alloys}

We will restrict the discussion mainly to the standard dilute alloy SGs, although the general conclusions can be applied to other families of Heisenberg SGs. It is important also to be able to make comparisons with results on an insulating Heisenberg SG, a diluted thiospinel  $CdCr_{2x}In{2(1-x)}S_4$ \cite{dupuis:02}, and with a canonical Ising SG material, Fe$_{0.5}$Mn$_{0.5}$TiO$_3$ \cite{ito:86,gunnarsson:88,gunnarsson:91}$^)$, which have been intensively studied experimentally .

Some dilute transition metal element impurities in noble metal hosts, in particular ${\bf Au}Mn$, ${\bf Ag}Mn$, ${\bf Cu}Mn$, ${\bf Au}Cr$, ${\bf Au}Fe$, have low or very low Kondo temperatures, so once the impurity concentration is not infinitessimal the interactions dominate over the Kondo effect and each individual impurity can be considered to be to a good approximation a classical Heisenberg local moment. There are $J_{ij}{\bf S}_{i} \cdot {\bf S}_{j}$ exchange interactions between the neighbors thorough the conduction electrons. Broadly speaking these interactions can be considered to be oscillating in sign (and so frustrated) with $<J_{ij}> \sim 0$ and "RKKY-like" in form even though the well known RKKY expression is a long distance approximation, and the effective interactions at distances up to a few atomic spacings deviate strongly from it.

It has been known empirically since the 1930s that these alloys are weakly magnetic with high temperature susceptibilities of the Curie-Weiss form $\chi(T) \sim 1/(T+\theta)$, where $\theta$ is a low Curie-Weiss temperature which can be either positive or negative and represents a deviation from the approximation that the average interaction $<J_{i,j}>$ is strictly zero. Ideally the impurities are distributed completely at random in the host matrix, but
there can be weak clustering or anti-clustering metallurgic effects (see Ref.~\cite{peil:09}$^)$ for an {\it ab initio} calculation of the metallurgical effects in ${\bf Cu}Mn$), which depend in some cases on the heat treatment given during the preparation of an individual sample.  The observed "spin glass cusp" temperature $T_c$ is well defined in dc field cooled (FC) low field susceptibility $\chi(T)$ measurements, while it is weakly frequency dependent in low field low frequency ac susceptibility measurements. It is accompanied by no sharp feature in the curves for the specific heat $C_v(T)$ or the resistivity $\rho(T)$. Irreversibility (classically, a difference between FC (field cooled) and ZFC (zero field cooled) susceptibilities) sets in exactly at $T_c$ in the low field limit.

\section{Dzyaloshinsky-Moriya anisotropy}

An important and subtle property specific to Heisenberg spin glasses (HSGs) is a particular form of Dzyaloshinsky-Moriya (DM) \cite{dzyaloshinsky:58,moriya:60}$^)$ anisotropy \cite{fert:80,fert:82}$^)$. Quite generally, in any system the DM interaction between two spins ${\bf S}_{i}$ and ${\bf S}_{j}$ is indirect and is through a spin-orbit coupling on a third site (which can be non-magnetic). Taking the third site as the origin, the interaction can be written in the form
\begin{equation}
H = - {\bf D}_{ij}\cdot[{\bf S}_{i} \times {\bf S}_{j}]
\end{equation}
where $ {\bf D}_{ij} = Q_{ij}( {\bf R}_{i}\times {\bf R}_{j})$, $Q_{ij}$ being the strength of the coupling, which is proportional to the spin-orbit interaction at the third site.

In a disordered HSG alloy the $D_{ij}$ vectors (each perpendicular to the plane containing the three sites considered) are distributed randomly in strength and in direction.  Each time a HSG sample is cooled below $T_c$
either in field or in zero field, the entire set of spins conspires to
minimize the total spin-spin interaction plus DM anisotropy
energy by taking up an appropriate configuration. In the "frozen" state there is then a built-in hidden global anisotropy ; each individual spin after each particular cooling cycle has a preferential local anisotropy axis in space, due to the DM interactions with all its neighbors in the particular complex configuration into which the system has been frozen. If the entire spin ensemble is then forced to turn rigidly through a small angle $\theta$ with respect to the lattice (by turning the applied field about any axis in the FC case and by the application and subsequent rotation of a magnetic field in the ZFC case) there is a directional anisotropy energy $E_{A} = K(T)(1-\cos(\theta))$ tending to return the spin system as close as possible to its initial position, because the spins prefer their original orientations in space. Suppose the spin glass has a strictly rigid spin configuration with magnetization $M$ and a field-independent
spin glass anisotropy $K$; then when the applied field $H$ is turned by an angle $\theta$ the torque signal $\Gamma(H)$ is given by \cite{fert:82}$^)$
\begin{equation}
\theta/\Gamma(H) = 1/K +1/MH
\end{equation}
For a series of points each taken after cooling in field, the torque signal $\Gamma(H)$ will initially increase with field as $H^2$ because the magnetization is proportional to $H$; when the limit $HM >> K$ is reached the torque will saturate at a field
independent value depending only on $K\sin(\theta)$ \cite{fert:82}$^)$. The anisotropy strength $K(T)$ can vary considerably from case to case; it depends on the intrinsic strength of the DM coupling for that particular alloy, on the relative temperature of the measurement (roughly as $K(T) \sim   K(0)(1-T/T_c)$), and weakly on time. The low temperature $K(0)$ limit is high for an alloy with strong spin-orbit coupling such as ${\bf Au}Fe$ \cite{petit:99}$^)$; in alloys such as ${\bf Cu}Mn$ where $K(0)$ is intrinsically weak \cite{petit:02}$^)$ it can be dramatically increased by the introduction of low concentrations of additional non-magnetic impurities having strong spin-orbit couplings \cite{courtenay:84,fert:85}$^)$.  Manifestations of the anisotropy include square hysteresis loops at temperatures $T << T_c$ \cite{kouvel:61,senoussi:85}$^)$, NMR effects \cite{alloul:80}$^)$ and above all magnetic torque \cite{fert:82,courtenay:84,fert:85,petit:99,petit:02,petit:02t}$^)$ which will be discussed in detail below.

\section{Spin Glass ordering and Critical exponents}

Present experiments cannot measure the spin glass $q_{SG}(T,H)$ or chiral glass $q_{CG}(T,H)$ parameters directly but only indirectly through the magnetization.
It is instructive to first follow the analysis of the simplest case of the EA Ising model.

In the symmetrical interaction distribution case, if the interaction between spins $i$ and $j$ is $J_{i,j}$, $<J_{i,j}>=0$ and the overall SG interaction strength parameter is the average $<J^2> = [J_{i,j}^2]_{av}$ over the interaction distribution  (rather than being $<J>$ as in the ferromagnetic case). From obvious dimensional considerations it is clear that that in SGs the "temperature" parameter must be $T^2$ and the "field" parameter $H^2$ (rather than $T$ and $H$ as in the ferromagnet case). For an ISG with a finite ordering temperature $T_{SG}^2$ is proportional to $<J^2>$.
Above $T_c$ each spin $i$ is subject to a random fluctuating field due to the interactions with its neighbors, a random variable whose variance is $<J^2>q$. In zero applied field the time average of this fluctuating field is zero, and the overall leading (linear) magnetic response to an infinitessimal magnetic field is the same as if there were no interactions, hence the linear term in the induced magnetization is "free-spin-like", $M(H,T) \sim H/T$.
The "reduced spin glass susceptibility" $\chi_{SG}(T)$ as measured conventionally in simulations represents the fluctuations in $q_{SG}(T,0)$
at zero applied field, $\chi_{SG}(T) = [<S_{i}\cdot S_{j}>_{T}^2]_{av}$.
Still above $T_c$, through the fluctuation-dissipation theorem the $q_{SG}(T,H)$ induced by a weak field is $q_{SG}(H,T) = (H^2/T^2)\chi_{SG}(T)$, so to next order through the spin-spin interactions there is an extra random static molecular field on each site $i$ due to this induced $q_{SG}(H,T)$ at the neighbors, and this affects $M(T,H)$. In the mean field case self consistent equations can be written down \cite{sherrington:75}$^)$ for $M(T,H)$ and $q_{SG}(H,T)$  which are exact for all $T$ above $T_c$.

Finally below $T_{SG}$ and in very small field
there are randomly distributed molecular fields which are acting on the individual spins and which no longer time average to zero due to the non-zero $q(T,0)$. The spins can no longer react to a small external field as if they were free-spin-like; in consequence under very low applied fields there is an abrupt change in the slope of $\chi(T)$ with temperature at $T_{SG}$. This is the cusp in the linear magnetic susceptibility.

In the mean field case (see Ref. ~\cite{bouchiat:86}$^)$), from the self-consistent SK equations
\begin{equation}
q(H,T) \sim (H^2/T^2)/(1-T_c^2/T^2)
\end{equation}
and so up to order $H^3$
\begin{eqnarray}
M(H,T) = H/T -(1/3)H^3/T^3 - (HJ^2/T^3)q_{SG}(H,T) \nonumber \\
\sim H/T -(1/3)H^3/T^3 - (H^3J^2/T^5)/\tau
\label{Mmf}
\end{eqnarray}
where $\tau =1-(T_{SG}/T)^2$.
The first two terms are the leading terms in the $S=1/2$ "free spin" magnetization $M(H,T) = \tanh(H/T)$ (it should be noted that there is a small "free spin" term in the non-linear magnetization response) and the third is the leading "spin glass" term.
The spin glass term diverges at $T_{SG}$, and below $T_{SG}$ finite spin glass order $q$ is frozen in. This result is exact for all $T$ above $T_{SG}$ in the mean field limit, in appropriately normalized units where $T_{SG} = J$; there are no corrections to scaling in mean field.

In dimensions below mean field \cite{daboul:04,campbell:06}$^)$ the reduced SG susceptibility is
\begin{equation}
\chi_{SG}(T) = C_{\chi}(\tau)^{-\gamma}[1 +a_{1}\tau^{\theta} + b_{1}\tau +...]
\label{wegner}
\end{equation}
where $\tau = 1-(T_{SG}/T)^2$, $\gamma$ is the susceptibility critical exponent and the small terms within $[1+...]$ are corrections to scaling (as in Ref.~\cite{wegner:72}$^)$ for the ferromagnetic case). The leading magnetization terms are then
\begin{eqnarray}
M(H,T)= H/T -(1/3)H^3/T^3 - (HJ^2/T^3)q_{SG}(H,T) \nonumber \\
\sim H/T -(1/3)H^3/T^3 - (H^3J^2/T^5)\tau^{-\gamma}[1+...]
\label{Mgamma}
\end{eqnarray}
Painstaking experimental studies (see in particular Refs.~\cite{miyako:80,monod:82,bouchiat:86,levy:88}$^)$) established the existence of and measured the critical divergence of the non-linear susceptibility thus demonstrating that the SG transition in laboratory samples is indeed a thermodynamic transition, albeit of an entirely novel type.

An analogous discussion provides a scaling rule one further step beyond beyond the small $H$ limit, which can used to extract the critical exponent $\beta$. The expression becomes
\begin{eqnarray}
M_{SG}(H,T)/(H^{2\beta/(\beta+\gamma)}T^{(3\gamma+5\beta)/(\gamma+\beta)}) \nonumber \\
 \sim G[(1-(T_{SG}/T)^2)/(H/T)^{2/(\beta+\gamma)}]
\label{Mbeta}
\end{eqnarray}
where $G[x]$ is a scaling function tending to a constant for small $x$ (high fields) and to $x^{-\gamma}$ for large $x$ (small fields). Here $M_{SG}(H,T)$ is the magnetization once an estimate for the "free spin" contribution (equal to $\tanh(H/T)$ in the spin $1/2$ Ising case) has been subtracted out. (This scaling rule is intrinsically invalid when $H$ becomes too large \cite{bouchiat:86}$^)$).

For historical reasons non-linear magnetization data have generally been analysed using \cite{suzuki:77,chalupa:77,geshwind:90}$^)$
\begin{equation}
M(H,T)- H/T \sim H^3(T-T_{SG})^{-\gamma}
\end{equation}
and
\begin{eqnarray}
(M(H,T)- H/T)/(H^{2\beta/(\beta+\gamma)}) \nonumber \\
\sim G[(T-T_{SG})/H^{2/(\beta+\gamma)}]
\end{eqnarray}
which are approximations valid in the limit $(T-T_{SG}) << T_{SG}$.
If measurements are not restricted to a very narrow range of temperature above $T_{SG}$ and sufficiently low applied fields the use of these simplified expression biases estimates of $\gamma$ and $\beta$, which explains certain apparent inconsistencies among published exponent estimates. The expressions $[\ref{Mgamma}]$ and $[\ref{Mbeta}]$ can extend the validity of the analyses much further in temperature and so give better based estimates for the exponents.

Exactly at $T_{SG}$ the leading critical behavior of the non-linear susceptibility is given by
\begin{equation}
M(H,T_{SG})- H/T_{SG} \sim H^{2/\delta}
\end{equation}
One has the scaling relation $\delta = [d+2-\eta]/[d-2+\eta]$, so a measurement of $\delta$ can be directly translated in terms of $\eta$ which is very convenient.

It should be noted that the exponent $\beta$ can be estimated independently from measurements below $T_{SG}$ ~\cite{bouchiat:86}$^)$. If the leading linear "free spin" susceptibility $\chi_{0}(H,T) \sim 1/T$ from above $T_{SG}$ is extrapolated to temperatures a little below $T_{SG}$, then from the discussion above, for small $H$ and $T < T_c$
\begin{equation}
\chi_{0}(H,T) - M_{FC}(H,T)/H \sim  q(T)/T^3 \sim ((T_{SG}/T)^2-1)^{\beta}/T^3
\label{Mbeta2}
\end{equation}
where $M_{FC}(T,H)$ is the field cooled magnetization.
Surprisingly this method, which should give accurate results as it is based on linear
susceptibility measurements, has been little used in practice.

All the preceding discussion concerns an ordering of the spin glass parameter $q_{SG}$ in the Ising context. In the case of vector spins, if the parameter which orders is the chirality, there will be a chiral $q_{CG}$ defined through the chiral variables Eqn. ~ \ref{chirality}. For the experimentally measured linear and non-linear susceptibilities a similar formal analysis as above can be expected to hold {\it mutatis mutandis} with $q_{CG}$ in the place of $q_{SG}$.

We can now review the experimental data for the critical exponents. From the discussion above it can be seen that the exponents the most readily measured experimentally are $\gamma$, $\delta$ and $\beta$. Using the standard scaling relations one can immediately derive estimates for the exponents $\eta$ and $\nu$, which are the exponents most directly obtained from analyses of numerical data : $\eta = 2-d(\delta-1)/(\delta+1)$, $\nu = \gamma/(2-\eta)$; $\beta = \gamma/(\delta-1)$ which provides a consistency check, and finally $\alpha = 2 - d\nu$. It turns out that $\alpha$ is always strongly negative in all 3d SGs, which means that there is no visible signature of the transition in specific heat measurements.

As a bench-mark, the exponents of the laboratory ISG material Fe$_{0.5}$Mn$_{0.5}$TiO$_{3}$ \cite{gunnarsson:91}$^)$ are shown in Table I where they are compared to simulation data for the 3d ISG with Gaussian near neighbor interactions \cite{katzgraber:06}$^)$. (Simulation values reported for the 3d ISG with binomial interactions are similar \cite{katzgraber:06,hasenbusch:08}$^)$ but this system is subject to strong finite size corrections to scaling \cite{hukushima:09}$^)$ so the values are less reliable). Globally, agreement between experiment and numerics is reasonably good; in particular both experiments and simulations indicate that $\gamma \sim 5$ which is high, and that $\eta \sim -0.35$ is strongly negative.

The Heisenberg SG which has been the most intensively studied is ${\bf Ag}Mn$ which together with ${\bf Cu}Mn$ is one of the canonical low anisotropy HSGs. Three independent sets of careful measurements \cite{bouchiat:86,levy:88,petit:02t}$^)$ used either dc or ac techniques and are entirely consistent with each other. In the former (see e.g. Ref. ~\cite{bouchiat:86}$^)$) field cooling measurements of the magnetization are made in a series of constant applied fields and the critical temperature $T_c$ is estimated from the position of the linear susceptibility cusp. The exponent $\gamma$ is estimated from the temperature dependence of the non-linear susceptibility term in $H^3$ above $T_c$, and the exponent $\delta$ from $M(H,T_c)$ at criticality. In the latter technique, weak ac fields are applied at a set of low frequencies $\omega$ and the strength of the harmonics at frequencies $3\omega, 5\omega,...$ are registered (see Ref. ~\cite{levy:88}$^)$). With this method the static $T_c$ must be estimated by extrapolating to zero frequency to allow for the intrinsic relaxation of the spin glass system just above $T_c$ as carefully explained in Refs. ~ \cite{gunnarsson:88,gunnarsson:91}$^)$. The exponents $\gamma$ and $\beta$ are estimated from the temperature dependence of the strengths of the harmonics near $T_c$. This method has the advantage of detecting directly and independently the non-linear terms in the susceptibility at
very low applied fields. The directly measured ${\bf Ag}Mn$ exponents are $\gamma = 2.1(1)$, and $\delta = 3.2(3)$ meaning $\eta = 0.4(1)$. The contrast with the ISG results is striking particularly for $\eta$ which instead of having a strongly negative value has a strongly positive value, while $\gamma$ is much lower in the HSG as compared to the ISG.

It is very instructive to compare these weak anisotropy limit experimental values with the most recent estimates from simulations for the critical exponents at the pure chiral ordering transition in the 3d HSG with Gaussian interactions, $\gamma = 2.0(4), \eta = 0.6(2)$ \cite{viet:09a}$^)$. (These estimates are in fact consistent with earlier numerical estimates for chiral exponents in HSGs \cite{hukushima:00,hukushima:05}$^)$). Thus the experimental exponent values for a low anisotropy HSG are very similar to the numerical values for the critical exponents at a pure chiral ordering transition, while they are dramatically different from the exponents (experimental or numerical) at a spin glass transition in an ISG.

It is important to note that ferromagnets show no such brutal variations of exponents with spin dimensionality $N$ within a given space dimension; in dimension 3 the  Ising ferromagnet ($N=1$) has the exponents $\gamma = 1.2372(1),\eta = 0.0368(1)$ \cite{deng:03}$^)$; the XY ferromagnet ($N=2$) has the exponents $\gamma = 1.3178(2),\eta = 0.0381(1)$ \cite{campostrini:06}$^)$,  while the Heisenberg ferromagnet ($N=3$) has the exponents $\gamma = 1.3960(1),\eta = 0.0375(1)$\cite{campostrini:02}$^)$. Thus though ferromagnets with different $N$ are not in the same universality class, in practice the differences between the actual values of the critical exponents ( particularly in the case of $\eta$ ) are minimal. In spin glasses not only are the measured values of the exponents for both ISGs and HSGs quite different from those of the the equivalent ferromagnets, but there are extremely strong changes in exponent values estimated from simulations when going from the ISG spin glass limit to the HSG chiral glass limit, and these differences are mirrored in the experimental data.

Table I also gives experimental estimates for a number of HSGs for which the DM anisotropy has been directly measured \cite{courtenay:84,vincent:87,petit:02t}$^)$. The exponents were obtained following the same experimental dc protocol \cite{bouchiat:86}$^)$ as in the case of the ${\bf Ag}Mn$, i.e. measurements were all made in a restricted range of field such that $\chi_{nl}(H) < 0.1\chi_{0}$, and within a restricted range of temperatures, $(T - T_c) < 0.1T_c$. These conditions ensure that the estimates for the different materials are as reliable as the results for ${\bf Ag}Mn$, and material-to-material comparisons are fully valid.
The experimental exponent estimates evolve systematically with the anisotropy strength, from close to the model chiral values for weak anisotropy to close to the model Ising values for strong anisotropy. There is no clear indication of a unique universality class for all HSG materials.

\begin{table}[htbp]
\caption{\label{Table:1} Values of the ratio of anisotropy constant (in erg/mol of magnetic sites) to $T_c$ \cite{petit:02}$^)$, and the critical exponents }
\begin{ruledtabular}
\begin{tabular}{cccccc}
sample
&$K(0)/T_c \times 10^{-5}$&$\eta$&$\nu$&$\gamma$&$\delta$ \\
&&&&&\\
Bimodal chiral\cite{hukushima:05}$^)$&simulation&0.8(2)&1.2(2)&1.5(4)&2.3(4)\\
Gaussian chiral\cite{viet:09a}$^)$&simulation&0.6(2)&1.4(2)&2.0(5)&2.75(4)\\
\textbf{Cu}Mn\cite{petit:02t}$^)$  &0.068&0.4(1)&1.3(1)&2.2&3.3\\
\textbf{Ag}Mn\cite{bouchiat:86}$^)$  &0.16&0.46(10)&1.40(16)&2.2(2)&3.1(2)  \\
\textbf{Ag}Mn\cite{levy:88}$^)$  &0.16&0.4(1)&1.30(15)&2.3(2)&3.3(3)  \\
CdCr$_2$InS$_4$\cite{vincent:87}$^)$ &0.8&0.17(10)&1.3(3)&2.3(4)&4.1(4) \\
\textbf{Au}Fe\cite{petit:02t}$^)$  &1.32&0.0(1)&1.6&3.2(3)&5.0(7) \\
(Fe$_{0.1}$Ni$_{0.9}$)$_{75}$P$_{16}$B$_6$Al$_3$\cite{petit:02t}$^)$ &2.65&-0.1(1)&1.7(2)&3.50(35)&5.7(8) \\
&&&&& \\
Fe$_{0.5}$Mn$_{0.5}$TiO$_3$\cite{gunnarsson:91}$^)$&ISG&-0.35(10)&1.7(3)&4.0(3)&8.4(15) \\
Gaussian ISG\cite{katzgraber:06}$^)$&simulation&-0.37(5)&2.44(9)&5.8(3)&8.5(8) \\
\end{tabular}
\end{ruledtabular}
\end{table}

\section{Torque measurements and in-field transitions}

While there have been numerous longitudinal field magnetization and susceptibility
measurements with increasingly subtle protocols on many SGs (see Ref.~ \cite{jonsson:04,vincent:06}$^)$), torque measurements after field turning can provide complementary information on irreversibility and relaxation, inaccessible to techniques where the field orientation is held fixed.

In a polycrystalline HSG the torque signal exists only because of the local random Dzyaloshinski-Moriya interaction discussed above. In particular the very existence of a SG torque signal after an applied field is rotated through a small angle implies that the global spin system has not attained equilibrium with respect to the new field direction, or in other words that the entire spin system is "frozen" and cannot reorganize microscopically so as to release the DM anisotropy constraint. The onset of in-field ordering as a function of field and temperature can be monitored by systematic torque measurements \cite{petit:99,petit:02,petit:02t}$^)$. An advantage is that rather than comparing FC and  ZFC magnetizations as in longitudinal irreversibility measurements, the technique compares an observed signal with a nul signal.

The principal experimental protocol consists in cooling a polycrystal SG sample in an applied field $H$ down to a temperature $T$. Once the temperature has stabilized, the field is turned through a small angle $\theta$ (typically $5$ degrees); the torque signal is then monitored continuously as a function of time. Experimental details are given in Ref.~ \cite{petit:02t}$^)$. Initial experiments were performed at temperatures well below $T_c$ in order to demonstrate the unusual unidirectional character of the SG anisotropy \cite{fert:82}$^)$. Experiments were later carried out in the region of $T_c$ up to fields of $7$ Tesla for a strong anisotropy ${\bf Au}Fe$ sample and to $4$ Tesla for various samples including weak anisotropy  ${\bf Cu}Mn$ and ${\bf Ag}Mn$ samples, the thiospinel CdCr$_2$InS$_4$, and the amorphous metal SG (Fe$_{0.1}$Ni$_{0.9}$)$_{75}$P$_{16}$B$_6$Al$_3$ \cite{petit:99,petit:02,petit:02t}$^)$. It should be kept in mind that for reasons of experimental sensitivity the effective applied transverse field $H\sin\theta$ was as high as $\sim 0.5$ Tesla, corresponding to a perturbation which is far from negligible. Measurements with an independent transverse field coil would be preferable but would require a different experimental set-up.

At temperatures above $T_c$ as determined by the linear susceptibility cusp no torque signal is observable on the time scale (a few seconds) of the measurement. On this time scale the spins have reorganized entirely in response to the new orientation of the field; there is no SG order. For temperatures somewhat below $T_c$ a finite torque signal persists over very long times even when the applied field is strong, demonstrating that SG order is present and is not destroyed by the field. In detail the behaviour depends drastically on the strength of the DM anisotropy. Concentrating first on the weak anisotropy SG ${\bf Ag}3\% Mn$ the relaxation of the torque signal, figure 1, is of the form $\Gamma(H,T,t) \sim t^{-\alpha(H,T)}$ within experimental accuracy right up to temperatures very close to $T_c$. This relaxation without any apparent characteristic time scale is precisely the functional form in the ordered state of an ISG or a HSG as observed in the low field limit for longitudinal magnetic measurements below $T_c$ (see e.g. Ref. ~\cite{ito:86}$^)$). However the contrast between the in-field behaviour of the ISG longitudinal relaxation and that of the HSG torque relaxation is striking. For the ISG  Fe$_{0.5}$Mn$_{0.5}$TiO$_3$ sample with $T_c \sim 20K$, at $T/T_c = 0.5$ under an applied field of 2 Tesla a characteristic decay time of $\tau \sim 60$ seconds has set in, and by 3 Tesla $\tau$ has dropped to $\sim 0.0006$ seconds \cite{mattsson:95}$^)$. These results were taken as evidence that there is no in-field transition in an ISG \cite{mattsson:95}$^)$, though the extrapolation to very low fields is delicate. In agreement with the experimental conclusions, numerical data indicate that the Ising spin glass transition does not survive under applied field \cite{jorg:08,katzgraber:09}$^)$ except possibly if these fields are very weak \cite{leuzzi:09}$^)$. For the HSG alloy ${\bf Ag}3\% Mn$ on the other hand with a rather lower transition temperature, at $T/T_c \sim 0.9$ (so much closer to $T_c$) and in fields of up to 4 Tesla (the limiting field of the experimental set-up) the torque signal decay has no apparent characteristic time and the decay exponent $\alpha(H,T)$ is practically field independent, figure 1. Na\"{i}vely one might expect that if the ordering mechanism was the same in HSGs as in ISGs, the latter would be much more robust against applied fields. In practice exactly the opposite is true.

\begin{figure}
\includegraphics[width=4in]{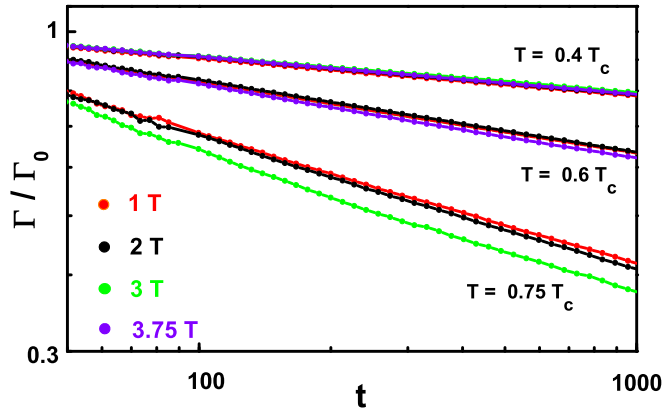}
\caption{(Color on line) The relaxation of the torque signal $\Gamma(t)$ normalized to the initial signal (10 seconds after turning), for ${\bf Ag}3\%Mn$. Applied fields are 1, 2 , 3 and 3.75 Tesla. The algebraic relaxation is field independent until close to $T_c$ where a weak field dependence of the relaxation exponent $\alpha$ sets in.}
\protect\label{fig:1}
\end{figure}

The frontier between the region of no observable torque signal and an observable finite torque signal with weak algebraic relaxation depends slightly on the experimental conditions and on the precise operational definition of an "unobservable" signal.   However using a consistent definition of the point at which the signal can no longer be observed, an effective $H(T)$ transition line can be defined and is shown for five HSGs  in figure 2. For the weak anisotropy alloy ${\bf Cu}3\% Mn$ this transition line rises almost vertically at a temperature $\approx 0.95T_c$. The phase transition for a model Heisenberg spin glass in which a weak anisotropy term was introduced was shown in Ref. ~\cite{imagawa:04}$^)$. The experimental and model behaviors are strikingly similar, including the temperature of the vertical line which is slightly lower than $T_c$ in both cases. This offset is linked to the non-zero anisotropy.

If the numerical calculations can be taken as a quantitative guide-line, the in-field chiral transition will only be suppressed by extremely high fields. For a sample with $T_c = 20K$ the numerical data in Ref.~ \cite{imagawa:04}$^)$ would correspond to a transition suppressed to zero temperature only under a field of $\sim 400$ Telsa. Because of experimental limitations the persistence of the chiral ordered state cannot be tested to fields anything like as high as this. However the data are consistent with the ordering transition surviving until very high fields.

It can be remarked that turning the field even through a small angle is a perturbation on the sample. From then on it is out of equilibrium and the relaxation of the torque signal is a response to this disequilibrium. In fact {\it a posteriori} it can be seen that a different experimental protocol would have been useful in order to pin down the onset temperature of the unperturbed torque signal.  The torque could have been measured in a first run turning the field just after cooling to $T$, and then again in a second independent run turning the field after a long waiting time at $T$. If the initial torque signals in the two runs were identical it could be concluded that there was no intrinsic relaxation of the unperturbed spin system (i.e. before field turning) during the waiting time after cooling to $T$. This procedure was followed in one early experiment \cite{giovannella:87}$^)$ (together with other complex protocols) but was not carried out systematically at temperatures close to $T_c$.

When measurements are made on samples with higher anisotropies including an amorphous metal SG using just the same protocol, the critical fields at which the torque signals becomes unobservable with the same criterion drop strongly and systematically with anisotropy strength, Figure 2. One can plausibly interpret this observation in terms of a crossover from chiral order dominated behavior at very low anisotropy to spin order dominated Ising-like behavior at high anisotropy. The high anisotropy situation is not strictly Ising as the local anisotropy axes have random orientations, but the measurements are entirely consistent with an "Ising" limit at very high anisotropy having no in-field spin glass order.

\begin{figure}
\includegraphics[width=4in]{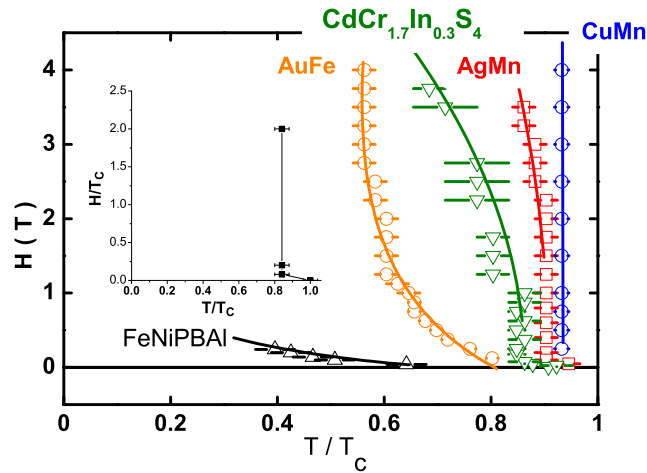}
\caption{(Color on line) The torque onset field (Teslas) against normalized temperature $T/T_c$ for five Heisenberg spin glasses : the alloys ${\bf Cu}3\%Mn$, ${\bf Ag}3\%Mn$, ${\bf Au}8\%Fe$, the thiospinel CdCr$_2$InS$_4$, and the amorphous metal (Fe$_{0.1}$Ni$_{0.9}$)$_{75}$P$_{16}$B$_6$Al$_3$. See Table I for the DM anisotropies. Inset : model $H(T)$ phase line for HSG with anisotropy $D=0.05$ \cite{imagawa:04}$^)$}
\protect\label{fig:2}
\end{figure}

The in-field onset of longitudinal irreversibility has also been studied carefully in HSGs (see e.g. Ref.~\cite{kenning:91}$^)$). A detailed comparison on the low anisotropy ${\bf Cu}Mn$ shows that at a given relative temperature even the "weak" irreversibility onset lies at a field significantly lower than that for the onset of a torque signal.

\section{Low field Dynamics and Memory effects}

The aging, memory and rejuvenation effects in spin glasses below $T_c$ are spectacular and have attracted considerable attention (see e.g  Ref. ~ \cite{jonsson:04,vincent:06}$^)$ ). For any SG, suppose the sample is cooled in a constant field down to temperature $T$ and is then kept at fixed $H$ and $T$ for a waiting time $t_w$. The field cooled magnetization $M(T,H,t_w)$ is always independent of $t_w$ to a very high approximation. However if the same cooling and waiting procedure is followed with $H=0$ and the ac susceptibility $\chi^{"}(T,t_w)$ is recorded, the ac susceptibility drops progressively with increasing time, a process known as aging. This shows that the internal state of the sample after the cooling procedure is not in true equilibrium despite the fact that the magnetization does not change with time once the temperature is fixed following a field cooled protocol. Temperature variation protocols have been used in very varied relaxation experiments \cite{dupuis:01,bert:04}$^)$.

Various complex protocols have been used to probe this remarkable behavior. For instance, a "memory dip" procedure can be followed in which the sample is cooled by stages with a number of stops at fixed temperatures during the cooling down. It can then be observed in a subsequent continuous slow heating that a "memory" of each stage has been imprinted in the sample ~\cite{jonason:98}$^)$. For present purposes let us concentrate on a simpler but instructive set of measurements \cite{dupuis:01,bert:04}$^)$. The sample is first cooled from above $T_c$ to a working temperature $T_1$ below $T_c$, and the ac  $\chi^{"}(T_1,t_w)$ is registered over a long time period $t_w = t_1 + t_2 + t_3$. In a second run exactly the same cooling procedure is followed, but in this run after waiting for time $t_1$ the temperature is reduced to $T_2 = T_1-\Delta T$ for a time $t_2$ (times used are typically $\sim 10^4$ seconds). The temperature is then increased back to $T_1$ and is held there for a further time $t_3$. During all this procedure $\chi^{"}(t)$ is registered. Immediately after the temperature has been decreased the initial $\chi^{"}$ is higher than before the temperature change; then while the temperature remains at $T_2$, $\chi^{"}(t)$ gradually decreases with time. However the most significant observation is what happens after the temperature returns to $T$. Comparing the two runs, after the $T_1/T_2/T_1$ procedure $\chi^{"}(t)$  with $t= t_1 +t_2 +t_3$ is the same as after the constant $T=T_1$ procedure with $t = t_1 + t_{eff} + t_3$. In other words, as far as the effective aging at $T_1$ is concerned, the time $t_2$ spent at $T_2$ is equivalent to having remained at $T_1$ for a time $t_{eff}$.

Results are strongly material dependent, and it has been observed that the DM anisotropy of the sample is an important parameter \cite{dupuis:01,bert:04}$^)$. It is instructive to compare data obtained for the two extreme cases : the Ising SG sample Fe$_{0.5}$Mn$_{0.5}$TiO$_3$ and the weak anisotropy Heisenberg SG ${\bf Ag}Mn$. Just below $T_c$, for the Ising sample $t_{eff}/t_2 \sim 1$ for temperature changes $\Delta T/T_1$ up to $5\%$, i.e. the time spent at $T_2 = 0.95T_1$ was almost as efficient for the $T_1$ aging of the sample as time spent at $T_1$. In contrast, for the ${\bf Ag}Mn$ sample just below $T_c$, much smaller temperature changes lead to drastically different results : for $\Delta T/T_1 \sim 1\%$, $t_{eff}/t_2 \sim 0$, i.e. time spent at $T_2 = 0.99T_1$  contributes almost nothing to the $T_1$ aging. Other Heisenberg samples with stronger anisotropies show behavior similar to but less extreme than that of the ${\bf Ag}Mn$ sample; however the sample-to-sample variation is clearly a function of the strength of the anisotropy \cite{bert:04}$^)$. These results from ac experiments are confirmed and reinforced by thermoremanent magnetization  relaxation experiments with similar temperature variation procedures \cite{bert:04}$^)$. It has been underlined that the multiple "memory" properties observed in multi-step cooling protocols are much sharper for low anisotropy Heisenberg SGs than for an Ising SG \cite{dupuis:01}$^)$. This property is directly linked to the results from the simpler $T_1/T_2/T_1$ protocol summarized above.

Interpretations of the rejuvenation and memory properties of laboratory spin glasses below $T_c$ have been given in terms of phase space and real space pictures (see Ref.~\cite{jonsson:04,vincent:06,vincent:09}$^)$). However no interpretation seems to have distinguished between dynamics under spin glass ordering and under chiral ordering. This type of analysis would appear to be well worth considering carefully in the light of the dramatic difference observed experimentally between the
Ising SG data and the weak anisotropy Heisenberg SG data. This difference suggests a chiral mechanism in the Heisenberg case. Kawamura and collaborators \cite{hukushima:00,imagawa:04,hukushima:05,viet:09}$^)$ have underlined the fact that in the simulations the chiral Binder parameter data, which go through a strong negative dip above $T_c$, should be interpreted in terms of a 1-step RSB transition. This corresponds to states below the transition temperature which are essentially orthogonal to each other. The "sharp memory" properties in low anisotropy HSGs may well be related to this particular characteristic of the HSG phase space \cite{kawamura}$^)$.

\section{Resistivity in mesoscopic samples}

Mesoscopic electrical resistance fluctuation experiments on metallic SGs \cite{weissman:98}$^)$ give insight into magnetic fluctuations at the level of $\sim 10^4$ spins. The resistivities of samples of lateral size $\sim 50$ nm or less are monitored as a function of time; the resistivity fluctuates because of the influence of the equilibrium magnetic fluctuations on universal conductance interference; universal conduction effects are well understood in similar sized non-magnetic metal samples. In the SGs ${\bf Cu}Mn$ and ${\bf Au}Fe$ there is a sharp rise in the $1/f$ noise in the resistivity when the temperature is lowered below $T_c$, and "non-monotonic wandering" of the spin noise is observed over very long times. The signal of the low anisotropy ${\bf Cu}Mn$ shows no sign of "droplet-like" excitations while the more anisotropic ${\bf Au}Fe$ has some weak "droplet-like" features \cite{weissman:98}$^)$. For this sort of measurement the samples must obviously be conducting, and it appears that there are no SG materials which are both Ising and conducting and which could provide a bench-mark to compare Heisenberg samples against. There are however metallic SGs with much stronger anisotropies than ${\bf Au}Fe$ and it might be useful to investigate them.

By sweeping the field applied to a mesoscopic sample and registering the variations in resistivity $\rho(H)$ linked to interference terms it is possible to obtain a "magnetic fingerprint" from which detailed information can in principle be deduced on the overlap properties of the sample \cite{vegvar:91}$^)$. At low temperature the magnetic structure of a ${\bf Ag}0.1\%Mn$ sample was found to be very robust against magnetic fields equivalent to $\sim 10T_c$ \cite{vegvar:91}$^)$. Again the 1-step RSB character of the chiral ordering may well be invoked, but for the moment it is premature to conclude that the behaviour observed is a specific signature of chiral ordering \cite{levy:09}$^)$.

\section{Hall effect}

In addition to the standard "Lorenz" Hall effect there is an "anomalous" Hall effect (AHE) associated with the magnetism in any magnetic conducting sample. This phenomenon has been known of for
well over a century, but only recently has there been a general consensus as to the major physical mechanisms leading to the effect (see Ref. ~\cite{onoda:09}$^)$ for a detailed overview). In the majority of cases the AHE is linked to the sample magnetization $M_z$ parallel to the applied field $H_z$, but when the spins of the sample are canted locally at random with respect to the z-axis (so that there are local spin components $S_{i,y}$ and $S_{i,x}$ ) it can be shown that there is an additional "real space Berry phase" contribution to the AHE which depends on the transverse $|S_{i,x}|$,$|S_{i,y}|$ \cite{tatara:02}$^)$. In a HSG under applied field, when there are frozen transverse components of the local moments an AHE term of this kind should be observed.

The analysis of measurements of the AHE in a series of
${\bf Au}Fe$ alloys demonstrates conclusively the presence of a
strong AHE a contribution linked to local spin canting in
addition to the standard intrinsic Kohn-Luttinger term \cite{pureur:04,wolff:06,taniguchi:04}$^)$. The results provide clear experimental evidence
which supports the theoretical predictions of a "chiral AHE" term
in disordered systems possessing chiralty. The strength of the effect predicted by theory is very delicate to estimate. The ${\bf Au}Fe$ alloy system turns out to be a
favorable case where the canting contribution dominates over much of the concentration range, probably because of the strong spin-orbit interaction. The chiral AHE can be understood physically in terms of a Hall current due to spontaneous
nanoscopic coherent current loops, a necessary consequence of time reversal
symmetry breaking in sequences of three or more scatterings
by tilted local spins. This mechanism has an entirely
different physical origin from that of the other contributions
which are invoked in interpretations of AHE data. Transport measurements are intrinsically "fast" and the chiral AHE can be expected to be strongly influenced by spin dynamics
through the coherence condition; the spin relaxation rate must be smaller than the
conduction electron scattering rate. Further work is needed to
better understood this effect quantitatively, so as to be able to obtain a direct measure of the chiral ordering.

\section{Conclusion}

The archetype laboratory dilute alloy spin glasses such as ${\bf Cu}Mn$, ${\bf Ag}Mn$ and ${\bf Au}Fe$ are Heisenberg; ever since the discovery of the spin glass phenomenon almost forty years ago there has been no consensus concerning the basic mechanism leading to the spin glass order in these materials (and in the vast majority of other laboratory spin glass systems, also HSGs). Chirality has been proposed \cite{kawamura:92}$^)$ as the driving order parameter in vector spin glasses (Heisenberg and XY). Simulations on these models are technically difficult because large sample sizes and rigorous equilibration at rather low temperatures are necessary, but
recent extensive numerical work on short range interaction Heisenberg spin glass models up to large sizes \cite{hukushima:05,viet:09a}$^)$ now appears to convincingly confirm the scenario \cite{kawamura:92}$^)$ in which chiral glass order sets in well before spin glass order, with a transition which is of 1-step RSB character.

It is important to examine the implications of the model analysis for the interpretation of the physical properties of the canonical laboratory spin glass systems. In real materials not only are the spins Heisenberg but the DM anisotropy is an important parameter. Luckily in ${\bf Cu}Mn$ and ${\bf Ag}Mn$ nature provides cases of systems with weak DM anisotropy; other materials such as ${\bf Au}Fe$ have considerably stronger anisotropies. The most direct tests of whether the chiral scenario is physically appropriate for interpreting the laboratory HSGs are comparisons between HSG model results and experimental measurements on HSG materials with weak anisotropy.

The critical exponents of these weak anisotropy alloy systems have been carefully measured \cite{bouchiat:86,levy:88,petit:02,petit:02t}$^)$; experimental exponents (typically $\eta \sim 0.5$ and $\gamma \sim 2.3$) are very similar to the HSG model values \cite{hukushima:05,viet:09a}$^)$($\eta = 0.6(2)$ and $\gamma = 2.0(4)$) and are strikingly different from numerical \cite{katzgraber:06}$^)$ or experimental \cite{gunnarsson:91}$^)$ Ising spin glass exponents, $\eta \sim -0.4$ and $\gamma \sim 5$. The measured values of the critical exponents for materials with stronger anisotropies lie between the model chiral spin glass values and the Ising values, tending progressively towards the Ising values with increasing anisotropy. There appears to be no sign of a unique HSG universality class.

Secondly, the model calculations show a chiral transition that is very robust under strong applied fields \cite{imagawa:04}$^)$; this can be intuitively understood in terms of  a weak coupling between field and chirality. Experimentally, torque measurements on the weak anisotropy materials show an effective in-field transverse order transition at a temperature close to $T_c$ which is almost independent of field up to the limit fixed by the experimental set-up. This is in complete contrast to the Ising case where the spin glass transition does not appear to survive under an applied field \cite{mattsson:95,jorg:08,katzgraber:09}$^)$. For HSG materials with stronger anisotropies the transverse order becomes progressively less robust against applied field, with a tendency towards "Ising-like" behaviour in the high anisotropy limit.

Thirdly, experiments show that memory and rejuvenation effects in the ordered state (under low fields) are strikingly different in the ISG and weak anisotropy HSG limits \cite{dupuis:01,bert:04,vincent:06}$^)$. A tentative interpretation of the HSG behaviour might link it to the 1-step RSB character of the chiral ordering shown by the numerical data on the model HSG \cite{kawamura}$^)$.

To summarize, the experimental data for weak anisotropy HSGs show numerous facets where behavior is entirely different from that of the Ising SGs. The chiral scenario provides a physical basis leading to a convincing interpretation of many of these HSG experiments. No alternative global interpretation of the experimental results has been proposed.

\section{Acknowledgements} We would like to thank Hikaru Kawamura, H\'el\`ene Bouchiat, Eric Vincent and Laurent L\'evy for their helpful remarks during the preparation of this manuscript.

\end{document}